\documentstyle[11pt]{article}
\begin{document}
\topmargin -1.4cm
\oddsidemargin -0.8cm
\evensidemargin -0.8cm
\topskip 9mm   
\headsep 9pt
%@twocolumnfalse\endcsname
\title{On Inverse Cascades and Primordial Magnetic Fields}
\author{\sc{P. Olesen$^1$}\\{\sl The Niels Bohr Institute, University
of Copenhagen}\\{\sl Blegdamsvej 17, DK-2100 Copenhagen \O\, Denmark}}
\maketitle
\vfill

\begin{abstract}
We consider primordial spectra with simple power behaviours and show that in
the Navier-Stokes and magnetohydrodynamics equations without forcing, there
exists systems in three dimensions with a subsequent inverse cascade,
transferring energy from small to large spatial scales. This can
have consequences in astrophysics for the evolution of density fluctuations,
for primordial magnetic fields, and for the effect of diffusion. In general,
if the initial spectrum is $k^{\alpha}$, then in the ``inertial'' range,
for $\alpha >-$3 there is an inverse cascade, whereas for $\alpha<-$3 there is
a forward cascade.
\end{abstract}
\vfill
\footnoterule
{\small $^1$Electronic address: polesen@nbi.dk}
\thispagestyle{empty}
\newpage
\setcounter{page}{1}
%%%%%%%%%%%%%%%%%%%%%%%%%%%%%%%%%%%%%%%%%%%%%%%%%%%%%%%%%%%%%%%%%%%%%%%%%%%%%%%

Recently there has been a considerable discussion of the development of
primordial magnetic fields subsequent to the creation of these fields (see
e.g. refs. \cite{silk}, \cite{brandenburg}, and \cite{b2}, where
further references can be found). In the early
universe the Reynold number $Re$ is often quite large. Since $Re$ is of the
order the ratio of the nonlinear to the linear terms, this suggest that the 
nonlinearities could be quite important. This was indeed found in ref.
\cite{brandenburg} and \cite{b2}, where an ``inverse cascade'' was observed.
In this note we point out that it is
an exact consequence of the Navier-Stokes and magnetohydrodynamics equations
without external forces that there exists an inverse cascade, moving
the field to larger scales, provided the initial energy is distributed at
relatively small scales. This result can be expressed as a scaling relation
for the relevant energy.  

In astrophysics one often considers ``primordial'' spectra behaving like
some power $k^{\alpha}$. The question we address in the following is
how such spectra evolve as consequence of the non-linear equations of
motion. We shall find that if the such a system is ``left alone'' (like in many
astrophysics applications), its energy is moved to larger and larger scales,
in contrast to a ``forced'' system (relevant for earthbound hydrodynamics) in
three dimensions, where energy is moved from larger to smaller scales. Hence
there is a fundamental difference between systems which are forced or left
alone.

In this note we shall consider hydrodynamics or magnetohydrodynamics (MHD).
In hydrodynamics one often considers systems which are under the
influence of forces, e.g. at large distances (``stirring''). In these systems
there is a forward cascade in three dimensions, which means that energy is
transferred from large scales to smaller scales (from ``order to chaos'').
As mentioned before, cases without forcing may also be of interest. In
particular, in astrophysics one may often encounter systems without any
essential ``stirring'', where the initial ``primordial'' spectrum is given. It
is then quite dangerous to apply results from forced systems. Here we shall
show that for a certain initial spectrum there exist an inverse cascade in
hydrodynamics (from ``chaos to order''), whereby
energy is transferred from smaller scales to larger scales, as a consequence
of the exact Navier-Stokes or magnetohydrodynamics (MHD) equations.

The most rigorous  of our results is that if the primordial energy spectrum is
given by $k$ at time 0, then at later times it will evolve as
\begin{equation}
k~\psi(k^2t),
\label{first}
\end{equation}
where $\psi$ is a scaling function (provided the viscosity coefficient is
constant in time) with $\psi(0)=1$. For MHD an equally rigorous result can be
derived for the magnetic energy. These results take into account diffusion. In
the ``inertial'' range, where diffusion can be ignored, a much more general
result, relevant for an initial spectrum $k^{\alpha}$, can be derived. In
order to have a finite energy, we must introduce an ultraviolet cutoff to
be specified more precisely later, so that $k\leq K$.
 
In a more special context the scaling behavior (\ref{first}) of the energy
density has been considered before, first by Heisenberg \cite{heisenberg}
in his effective diffusion model,
and then by Parisi \cite{parisi} in a continuous version of the cascade (GOY)
model \cite{goy}. A generalization of the scaling in (\ref{first}) was then
found in an investigation of the inverse cascade in the continuous GOY version
of three dimensional  relativistic MHD by Brandenburg {\it et al.}
\cite{b2}. Later the scaling (\ref{first})
has also been generalized to turbulent mixtures \cite{mp}. In the following
the result (\ref{first}) and its generalization to MHD are derived from the
exact Navier-Stokes and MHD equations, showing that Eq. (\ref{first}) is
much more general than previously thought \footnote{This result can be used 
to check if model calculations are in agreement
with the Navier-Stokes equation: If forcing is taken to be absent, the
scaling (\ref{first}) should be valid. From this point of view the cascade
(GOY) models are excellent, since they have the scaling as an exact solution
in the force-free case, as shown in Refs. \cite{parisi}, \cite{brandenburg},
and \cite{mp}.}. 

We start from the expression for the energy in $k=|{\bf k}|$-space,
\begin{equation}
E(k,t,L,K)=\frac{2\pi k^2}{(2\pi)^3}\int_{2\pi/K}^L d^3x d^3y~ 
e^{i{\bf k}({\bf x}-{\bf y})}~ <{\bf v}({\bf x},t)~
{\bf v}({\bf y},t)>,
\label{a1}
\end{equation}
with the total energy given by
\begin{equation}
\int_{2\pi/L}^K dk~E(k,t,L,K)=\int_{2\pi/K}^L d^3x~\frac{1}{2}
<{\bf v}({\bf x},t)^2>,
\label{bt}
\end{equation}
where $L$ indicates the insertion of an infrared cutoff to be specified 
precisely later.

Our initial spectra are taken to be simple powers. Thus, for powers
larger than or equal -1, we need the ultraviolet cutoff $K$ mentioned before.
If this is translated to the behaviour of the velocity correlator in
the integrand on the right hand side of Eq. (\ref{a1}), we need a short
distance cutoff $1/K$ in ${\bf x}$-space at time $t=0$. This is because 
the primordial spectrum is generated by some physical mechanism, and hence
the initial energy must be finite. Therefore $K$ is determined by the physical
mechanism which generates the initial spectrum.  For simplicity we assume
that the cutoff can be represented by a factor\footnote{For example, if
$F(x)=e^{-x}$, then this can be achieved in ${\bf x}$-space
by replacing some factors of the type 1/$({\bf x}-{\bf y})^2$ by combinations
of factors the type 1/$(|{\bf x}-{\bf y}|\pm i/K)$.}.
\begin{eqnarray}
F(k/K),&~&~{\rm where}~~F(x)\approx 1~~{\rm for}~~x\leq 1,\nonumber \\&~&
{\rm and}~F(x)\approx 0~~{\rm for}~~x\geq 1,
\label{skaer}
\end{eqnarray}
for $t=0$ in the primordial energy $E$. When $t$ increases, the initial
cutoff is no longer needed because of diffusion, from which we expect the
energy to decrease like exp$(-2\nu k^2t)$ for large values of $k$.

Next we want to use the well known self-similarity property of
the non-relativistic Navier-Stokes or MHD-equations,
\begin{equation}
{\bf x}\rightarrow l{\bf x},~~t\rightarrow l^{1-h}t,~~
{\bf v}\nonumber \\\rightarrow
l^h~{\bf v},~~\nu\rightarrow l^{1+h}\nu,~~{\bf B}\rightarrow
l^h ~{\bf B},~\eta\rightarrow l^{1+h}\eta,
\label{a2}
\end{equation}
where $\nu$ is the kinetic and $\eta$ is the Ohmic diffusion. Using the
substitutions ${\bf x}= l{\bf x}'$ and ${\bf y} =l{\bf y}'$, we obtain from
eqs. (\ref{a1}) and (\ref{a2})
\begin{eqnarray}
E(k/l,l^{1-h}t,Ll,K/l)&=&l^4\frac{2\pi k^2}{(2\pi)^3}\int_{2\pi/K}^{L} d^3x'
d^3y'~e^{i{\bf k}({\bf x}'-{\bf y}')}~<{\bf v}
(l{\bf x}',l^{1-h}t)~{\bf v}(l{\bf y}',l^{1-h}t)>\nonumber \\
&=&l^{4+2h}~\tilde{E}(k,t,L,K).
\label{a3}
\end{eqnarray}
Here $\tilde{E}$ deviates from $E$ in the sense that it is the energy computed
from a theory where $\nu$ is rescaled according to (\ref{a2}). The cutoff $K$,
introduced  by (\ref{skaer}), has the property from (\ref{skaer}) that it 
only depend on $k/K$, and hence is the same factor on both sides of Eq.
(\ref{a3}). For $t>0$ viscosity provides the cutoff by the factor
exp$(-2\nu k^2t)$ in the energy. However, the quantity $\nu k^2t$ is invariant
under the self-similarity transformation (\ref{a2}), and hence the diffusive
cutoff is also the same on both sides of Eq. (\ref{a3}). Thus, for $k\leq K$
we can leave out the $K$-dependence on both sides of Eq. (\ref{a3}), and in
the following we therefore leave out the argument ``$K$'' in the energy, and
we tacitly assume that $k\leq K$ in the following.

Let us define as usual the ``inertial range'' as the interval of $k$-values
where the dynamics is independent of the scale of dissipation, so that in this
range diffusivity can be taken to be zero. We assume that such a range of
$k$-values actually exists. If one of the conditions
\begin{enumerate}
\item[(i)] $h=-1$
\item[(ii)] $h\neq -1$, but we are in the ``inertial'' range,
\end{enumerate}
is satisfied, we have $E(k,t,L)=\tilde{E}(k,t,L)$. This is because if (i) is
satisfied, diffusion is left invariant in the self-similarity (\ref{a2}), and
if (ii) is satisfied, we can take $\nu$=0. In the following it is
assumed that one of these conditions are satisfied, and we therefore take
$E=\tilde{E}$. Eq. (\ref{a3}) then becomes a functional relation for the
energy, which we shall solve in the following. Introducing the energy density
$\cal E$ by $E={\cal E} V$, where $V$ is the normalization volume, defined
with the infrared cutoff as
\begin{equation}
V=\int_{-L}^{L}d^3x, 
\label{1}
\end{equation}
we have
\begin{equation}
{\cal E}(k/l,l^{1-h}t,lL)=l^{1+2h}~{\cal E}(k,t,L).
\label{a4}
\end{equation}
If the infrared cutoff is implemented in ${\bf x}$-space by an exponential
exp$(-|{\bf x}-{\bf y}|/L)$ in the integrals in Eqs. (\ref{a1}) 
and (\ref{a3}), it will appear in $k$-space by the substitution
$k^2\rightarrow k^2+1/L^2$ for some factors of $k$ in the spectrum ${\cal E}$.
This is because the exponential cutoff, after performing the angular
integration, is equivalent to the substitution
$|{\bf k}|\rightarrow |{\bf k}|\pm i/L$ in the integrals in (\ref{a1}) and
(\ref{a3}). For the simple primordial power spectra used in this note, it
therefore follows that as far as ${\cal E}$
is concerned, the limit $L\rightarrow \infty$ can be performed without
encountering any singularities except possibly in the point $k=0$, i.e.
\begin{equation}
{\cal E}(k,t)=\lim_{L\rightarrow\infty}~{\cal E}(k,t,L)
\label{lim}
\end{equation}
exists. Hence the cutoff dependence in Eq. (\ref{a4}) can be ignored in the
following.

The exception to Eq. (\ref{lim}) is that if ${\cal E}$ is formally singular in
$k=0$. In order to have the physically required finite energy, $k$ should be
substituted by $\sqrt{k^2+1/L^2}$. The value of $L$ should then be determined
by the physics which generates the primordial energy.

If we introduce the function $\psi$
\begin{equation}
{\cal E}(k,t)=k^{-1-2h}\psi(k,t),
\label{a5}
\end{equation}
we see that 
\begin{equation}
\psi(k/l,l^{1-h}t)=\psi(k,t).
\label{a6}
\end{equation}
Differentiating (\ref{a6}) with respect to $l$ and putting $l$=1 afterwards, we
get
\begin{equation}
-k{\partial \psi\over\partial k}+(1-h)t{\partial \psi\over\partial t}=0,
\label{a7}
\end{equation}
with the solution $\psi=f((1-h)\log k+\log t)$, where $f$ is an arbitrary
function. Thus, using eq. (\ref{a5}) we have the scaling relation 
\begin{equation}
{\cal E}(k,t)=k^{-1-2h}~\psi(k^{1-h}t).
\label{a8}
\end{equation}
As is clear from the preceeding discussion, this relation is valid if either
$h=-$1, or, more generally, if we are in the inertial range where diffusion
can be ignored. 

The general interpretation of (\ref{a8}) is that if initially the
``primordial'' spectrum is given by 
\begin{equation}
{\cal E}(k,0)=k^{\alpha},
\label{a9}
\end{equation}
then at later times the spectrum in the inertial range is given by
\begin{equation}
{\cal E}(k,t)=k^{\alpha}~\psi(k^{(3+\alpha)/2}t).
\label{a10}
\end{equation}
Depending on the value of $\alpha$ we may need an infrared cutoff.
If $h=-$1, we see from the self-similarity (\ref{a2}) that if the initial
spectrum has $\alpha$=1 ($h=-$1), then the spectrum is always given by
eq. (\ref{a10}), because the diffusion coefficient is the same in the scaled
and unscaled equations, i.e. 
\begin{equation}
{\cal E}(k,t)=k~\psi(k^2t)
\label{a11}
\end{equation}
{\it holds as a consequence of the Navier-Stokes equations for all values of
k}. It should be noticed that the reason this type of scaling works also when
diffusion is included, is that diffusion scales exactly like in
Eq. (\ref{a11}). The new thing is that the non-linear effects scale the same
way, if the initial spectrum is linear in $k$. Thus we see an inverse cascade,
instead of a cascade\footnote{It goes without saying that if $\psi$ is
dominated by diffusion, the result is
not interesting. Therefore the Reynold number should be so large that
non-linear terms are important.}.

In the case of MHD, Eq. (\ref{a11}) is still valid, and is to be
supplemented by the relation
\begin{equation}
{\cal E}_B(k,t)=k~\phi(k^2t),
\label{a12}
\end{equation}
where ${\cal E}_B$ is the magnetic energy density,
\begin{equation}
{\cal E}_B(k,t)=\frac{2\pi k^2}{(2\pi)^3V}\int d^3x d^3y~
e^{i{\bf k}({\bf x}-{\bf y})}~
 <{\bf B}({\bf x},t){\bf B}({\bf y},t)>,
\label{a13}
\end{equation}
and $\phi$ is another scaling function. We also used that ${\bf B}$ scales the
same way as ${\bf v}$, and the Ohmic diffusion scales like $\nu$.  It is now
quite clear that the spectrum has an inverse cascade: The
spectra start at $t$=0 with the linear form ${\cal E},{\cal E}_B\propto k$,
and then the dynamics at later times is governed by the scaling variable
$k^2t$, which is invariant for $k^2\propto 1/t$, so at later times, for
a given value of $\psi$, $k$ diminishes, and energy is transferred from
smaller to larger scales. 

From Eq. (\ref{a10}) we see that in general in the inertial range there is an
inverse cascade as long as $\alpha>-$3, otherwise we have a direct
cascade. This implies that if a (magneto-)hydrodynamical system is left to
itself, the way in which it develops is governed by the initial distribution of
the energy (for $k\leq K$): If much initial energy is concentrated at large
(small) distances, the system will cascade towards small (large) distances.

So far, we have considered the non-relativistic (magneto-)hydrodynamics
equations. The general relativistic case for an expanding, flat universe has
been discussed in ref. \cite{brandenburg}. The result is that the equations have
the same form as in a non-expanding universe, provided we scale the
energy density by $R^4$ and the magnetic field by $R^2$, where $R$ is the
scale factor. Furthermore, the dynamical time now becomes {\it conformal}
time $\tilde{t}$, defined by
\begin{equation}
\tilde{t}=\int^{t} dt'/R(t')\propto R(t)\propto \sqrt{t}.
\label{conf}
\end{equation}
Here $t$ is the Hubble time. The dynamical time $\tilde{t}$ is thus slowed down
relative to the Hubble time.
The velocity is not scaled, and the coordinates entering differentiation
are the comoving ones. Therefore, $k$ is the comoving wave vector. The
conserved energy is the energy multiplied by $R^4$. Therefore Eq. (\ref{a10})
is changed to
\begin{equation}
{\cal E}(k,\tilde{t})R(t)^4=k^\alpha~\psi(k^{(3+\alpha)/2}\tilde{t})
=k^\alpha~\tilde{\psi}(R(t)k^{(3+\alpha)/2}).
\label{11}
\end{equation}
Also, the magnetic energy is modified to
\begin{equation}
{\cal E}_B(k,\tilde{t}) R(t)^4=k^\alpha~\phi(k^{(3+\alpha)/2}\tilde{t})
=k^\alpha~\tilde{\phi}(R(t)k^{(3+\alpha)/2}).
\label{expand}
\end{equation}
These expressions are valid if assumption (ii) is satisfied, i.e. in the
inertial range where one can ignore diffusion. In most cases of interest in
astrophysics, diffusion is not constant in time. For example, for Silk
damping \cite{silk} the effective diffusion behaves as $\nu/R\propto
\tilde{t}~^2$. In such a case we can only say that the inverse cascade competes
with diffusion. Detailed calculations at high Reynold numbers are needed
in order to see whether the inverse cascade wins over the diffusion (see Ref.
\cite{b2}). It should be noticed that the results for the expanding
universe are valid in comoving coordinates. The expansion of the physical
coordinates, or the red shift of $k$, comes on top of the inverse cascade.

The fact that a (magneto-) hydrodynamic system behaves in a way which is
fundamentally different when it is forced or left alone, is perhaps not of much
interest in the usual earthbound hydrodynamics, since systems without forcing
die out. However, it can have consequences for astrophysics, because the
various spectra, e.g. density fluctuations, observed now may have been
subject to inverse cascades earlier. Thus, going backwards in time, the
primordial scales may have been smaller than what is generally thought from
data obtained now. This is especially true since in general the relevant
Reynold numbers are quite large in cosmology, so the non-linear terms are
expected to be important. Of course, in order to get quantitative estimates,
it is necessary to do explicit calculations in order to determine the spectra,
or equivalently, the scaling functions $\psi$ and $\phi$. This was already done
for the magnetic energy spectra by Brandenburg {\it et al.} in a shell model
for relativistic MHD in a flat, expanding universe, see Refs.
\cite{brandenburg} and \cite{b2}. Depending on the actual magnitude of the
inverse cascade, it may, for example, imply that less inflation is needed.

An inverse cascade has the effect of generating order from chaos. This was
clearly seen in Ref. \cite{b2}. To give an example, in the
case where the initial spectrum is linear in $k$, this spectrum corresponds to
initial short range velocity correlations in ${\bf x}$-space. However, as time
passes, the correlation gets a longer range, because $k\propto 1/\sqrt{t}$,
thereby creating more order.

In the previous discussion we have ignored the effect of gravity (except for
the inclusion of the scale factor). Since gravity is an attractive force, it
will transfer energy from larger to smaller scales, and hence it
will compete with the inverse cascade. In the non-relativistic case there is,
however, one case where there is still scaling with gravity included, namely
when $h=-1/2$. In this special
case the gravitational force scales like the other terms in the Navier-Stokes
equation. For $h=-1/2$ one has the initial energy ${\cal E}(k,0)=$const. In
this case, at later times one has ${\cal E}(k,t)=\psi(k^{3/2}t)$ in the
inertial range.     

Finally, inverse cascades have the general effect of diminishing diffusivity,
because energy is transferred away from the small scale region, where
diffusion is operative. Therefore, there will be more energy at larger scales
than one might infer by ignoring the non-linear terms in the equations of
motion. For MHD, the diminished effect of Silk damping \cite{silk} was seen
in Ref. \cite{b2}.   

I thank Larry McLerran for his continuous interest and encouragement, and
Mogens H\o gh Jensen for interesting discussions of hydrodynamics.


\begin{thebibliography}{X}

\bibitem{silk}
J. Silk, Astrophys.\ J.\ {\bf 151}, 459 (1968); K. Jedamzik, V. Katalinic, and
A. Olinto, preprint astro-ph/9606080.


\bibitem{brandenburg}
A. Brandenburg, K. Enqvist, and P. Olesen, Phys. Rev. {\bf D54}, 1291 (1996).


\bibitem{b2}
A. Brandenburg, K. Enqvist, and P. Olesen, preprint hep-ph/9608422 (1996),
Phys. Lett. B to be published.

\bibitem{heisenberg}
W. Heisenberg, Proc.\ Roy.\ Soc.\ A {\bf 195}, 402 (1948).

\bibitem{parisi}
G. Parisi, preprint of the University of Rome, unpublished (1990).

\bibitem{goy}
A. M. Obukhov, Atmos.\ Oceanic.\ Phys.\ {\bf 7}, 41 (1971); E. B. Gledzer,
Sov.\ Phys.\ Dokl.\ {\bf 18}, 216 (1973); V. N. Desnyansky and E. A.
Novikov, Prinkl.\ Mat.\ Mekh.\ {\bf 38}, 507 (1974);
M. Yamada and K. Ohkitani,  J.\ Phys. \ Soc. \ Japan \ {\bf 56}, 4210 (1987);
M. Yamada and K. Ohkitani, Prog.\ Theor.\ Phys.\ {\bf 79},
1265 (1988); M. H. Jensen, G. Palatin, and A. Vulpiani, Phys.\ Rev.\ {\bf A43},
798 (1991); {\it ibid} {\bf A45}, 7214 (1992); R. Benzi, L. Biferale, and
G. Parisi, Physica D {\bf 65}, 163 (1993); L. Kadanoff, D. Lohse, J. Wang,
and R. Benzi, Phys.\ Fluids \ {\bf 7}, 617 (1995); L. Kadanoff, Phys.\ Today \
{\bf 48}, 11 (1995); L. Biferale, A. Lambert, R. Lima, and G. Paladin,
Physica D {\bf 80}, 105 (1995); R .Benzi, L. Biferale, R. Kerr, and
E. Trovatore, Phys.\ Rev. \ {\bf E53}, 3541 (1996).

\bibitem{mp}
M. H. Jensen and P. Olesen, preprint cond-mat/9610047 (1996).


\end{thebibliography}
\end{document}